\documentclass[conference]{IEEEtran}
\IEEEoverridecommandlockouts
\usepackage{cite}
\usepackage{amsmath,amssymb,amsfonts}
\usepackage{algorithmic}
\usepackage{graphicx}
\usepackage{textcomp}
\usepackage{xcolor}

\usepackage[T1]{fontenc}    
\usepackage[utf8]{inputenc}  
\usepackage{tikz} 

\newcommand{\mycopyrightnotice}{%
  \begin{tikzpicture}[remember picture,overlay]
    \node[anchor=south, yshift=0.4in] at (current page.south) {%
      \parbox{\textwidth}{\fontsize{7.5pt}{9pt}\selectfont
      \copyright~2026 IEEE. Personal use of this material is permitted. Permission from IEEE must be obtained for all other uses, in any current or future media, including reprinting/republishing this material for advertising or promotional purposes, creating new collective works, for resale or redistribution to servers or lists, or reuse of any copyrighted component of this work in other works.\\
      Accepted for publication in the Proc. 48th Annu. Int. Conf. IEEE EMBS (EMBC 2026), Toronto, Canada, July 20-24, 2026.}
    };
  \end{tikzpicture}%
}

\def\BibTeX{{\rm B\kern-.05em{\sc i\kern-.025em b}\kern-.08em
    T\kern-.1667em\lower.7ex\hbox{E}\kern-.125emX}}
\begin{document}

\title{
Sleep EEG Signal Criticality as a Non-Invasive Predictor of Cognitive Decline in Dementia
}

\author{\IEEEauthorblockN{Stanisław Narębski$^\ddag $}
\IEEEauthorblockA{\textit{Academia Copernicana} \\
\textit{Nicolaus Copernicus University}\\
Toruń, Poland \\
ORCID: 0009-0004-8680-9499}
\and
\IEEEauthorblockN{Tomasz Komendziński}
\IEEEauthorblockA{\textit{Department of Cognitive Science} \\
\textit{Nicolaus Copernicus University}\\
Toruń, Poland \\
ORCID: 0000-0003-3273-0764}
\and
\IEEEauthorblockN{Tomasz M. Rutkowski$^\ddag $}\thanks{$^\ddag$Corresponding author.}
\IEEEauthorblockA{
\textit{The University of Tokyo}, Tokyo, Japan \\
\textit{Nicolaus Copernicus University}, Toruń, Poland \\
\textit{Araya Inc.}, Tokyo, Japan \\
ORCID: 0000-0002-4259-4121}
}

\maketitle

\mycopyrightnotice 

\begin{abstract}
Early detection of neurodegeneration remains a critical clinical challenge. This study investigates whether sleep EEG signal criticality, quantified via Multifractal Detrended Fluctuation Analysis (MFDFA), serves as a non-invasive biomarker for future cognitive decline. We analyzed longitudinal data from the National Sleep Research Resource (NSRR) Study of Osteoporotic Fractures (SOF) cohort, comparing baseline sleep EEG dynamics between women who remained cognitively normal and those who later progressed to dementia-related impairment ($3MS < 78$).Our results reveal significant group-level differences in Hurst exponent $H(q)$ distributions, particularly during non-REM stages N2 and N3. Cognitively healthy individuals exhibited signal dynamics significantly closer to an optimally critical state across all electrode locations ($p \leqslant 0.001$), supporting the Brain Criticality Hypothesis. Supervised UMAP projections confirmed clear spatial separation between groups throughout the overnight sleep architecture.The dementia group demonstrated a shift in DFA exponents toward $1.0$, suggesting that a reconfiguration of scale-free neural dynamics during sleep precedes clinical symptoms. These findings highlight the potential for MFDFA-derived measures to be integrated into automated, sleep-based screening tools, enabling earlier preventative interventions during the prodromal window of dementia.
\end{abstract}


\section{Introduction}

Characterized by a diverse spectrum of neurodegenerative pathologies, dementia represents one of the most pressing public health challenges of the modern era, most frequently manifesting as Alzheimer’s disease (AD)~\cite{livingston2024dementia}. Advanced age remains the most significant predictor of the condition, which is clinically defined by a loss of cognitive ability severe enough to disrupt independent daily living~\cite{livingston2024dementia}. The global burden is staggering: prevalence currently exceeds 57 million individuals, a figure anticipated to nearly triple to 150 million by mid-century as the demographic shift toward an aging population accelerates~\cite{livingston2024dementia}.

However, the trajectory toward dementia is rarely abrupt. Individuals who exhibit objective deficits in memory or executive function—yet maintain relatively preserved functional independence—are often classified as having Mild Cognitive Impairment (MCI)~\cite{petersen2008mild}. Within a five-year window following an MCI diagnosis, the probability of progressing to clinical dementia increases substantially~\cite{tuokko2003five}. Consequently, this intermediate phase offers a critical diagnostic "window of opportunity" to implement proactive interventions designed to prevent or delay further neurological decline. Given the significant risk of progression, there is an urgent need for reliable, non-invasive biomarkers that can detect subtle neurobiological changes before irreversible damage occurs.

Objective sleep metrics, such as those derived from polysomnography (PSG), represent a promising frontier for such early-stage identification and risk mitigation. Research suggests a U-shaped association between sleep length and neurodegenerative risk; both short sleepers and those with prolonged sleep durations exhibit higher rates of cognitive impairment~\cite{xu2020sleep, sabia2021duration}. Beyond duration, the micro-architecture of sleep plays a protective role against neurodegeneration. This is particularly evident in the structural organization of sleep stages and the proportion of time spent in deep non-REM sleep. Slow-wave sleep (SWS), characterized by high-amplitude delta oscillations ($<4$ Hz), represents a critical period during which the glymphatic system becomes active. This macroscopic waste clearance system eliminates metabolic byproducts—including neurotoxic proteins such as tau and $\beta$-amyloid—from the brain's interstitial space~\cite{xie2013sleep}. Disruptions in these oscillatory patterns may therefore serve as an early warning sign of accumulation pathology~\cite{haghayegh2025predicting}.

To detect these subtle disruptions, traditional linear methods may prove insufficient for analyzing complex EEG signals. 
Emerging research points toward non-linear dynamics, specifically the Brain Criticality Hypothesis, as a robust framework for understanding neural health. According to this hypothesis, neural architectures and their resulting functional patterns spontaneously self-organize into a critical state~\cite{wilting201925crit}.

The concept of criticality suggests that the healthy human brain operates at a phase transition point between order (rigid synchronization) and disorder (noise). This metastable state facilitates maximal information processing, heightened stimulus sensitivity, and flexible state transitions.

This "critical" state is characterized by long-range temporal correlations (LRTC) and self-similar (fractal) dynamics, where the fluctuations of neural signals exhibit a scale-invariant structure~\cite{tomen2019functional}. We hypothesize that neurodegeneration precipitates a deviation from this optimal critical regime, resulting in a loss of signal complexity that can be detected via EEG long before behavioral symptoms manifest. This critical state can be quantified using the Hurst exponent $H$, where values approaching $H \approx 1.0$ indicate the presence of scale-free, critical dynamics characteristic of healthy brain function~\cite{blythe2014effect}.

This paper presents a novel approach towards dementia prediction from sleep EEG by quantifying signal criticality measures. By analyzing the breakdown of scale-free dynamics during sleep, we aim to provide a sensitive biomarker for the transition from MCI to dementia.

\section{Methods}\label{sec:methods}

Data for this research were obtained with permission from the National Sleep Research Resource (NSRR)~\cite{zhang2018national}, utilizing a subset~\cite{spira2008sleep} of the Study of Osteoporotic Fractures (SOF) database~\cite{cummings1990appendicular}. The SOF study was a prospective US-based multicenter study that recruited women aged 65 years and older. In-home overnight polysomnography (PSG) was performed using a Compumedics P-Series system, recording four EEG channels on 461 participants. To focus on a cognitively normal group at baseline, the current analysis was restricted to participants with a Mini-Mental State Examination (MMSE)~\cite{folstein1975mini} score $> 24$ at the time of PSG who also completed a follow-up examination five years later ($N=290$). Participants were categorized into two groups based on their performance on the Teng-modified Mini-Mental State Examination (3MS)~\cite{teng1987modified} during the follow-up assessment. Subjects achieving a score of $78$ or higher were classified as cognitively normal, while those scoring below $78$ were identified as having cognitive impairment. The PSG recording montage included channels C3, C4, A1, and A2 (central and mastoid locations) sampled at $128$~Hz. 
The resulting data were segmented into 30-second epochs and manually scored by trained personnel into stages of wakefulness, rapid eye movement (REM) sleep, or non-REM stages (N1, N2, N3 and N4); for the present analysis, N3 and N4 from the original SOF records were consolidated into a single N3 stage to align with current clinical guidelines~\cite{iber2007aasm}.
Data post-processing and analysis were conducted under the approval of the Nicolaus Copernicus University in Toruń, Poland, Faculty of Philosophy Research Ethics Committee (Decision No. 60/2025).
\begin{figure*}[t]
\centerline{\includegraphics[width=\textwidth]{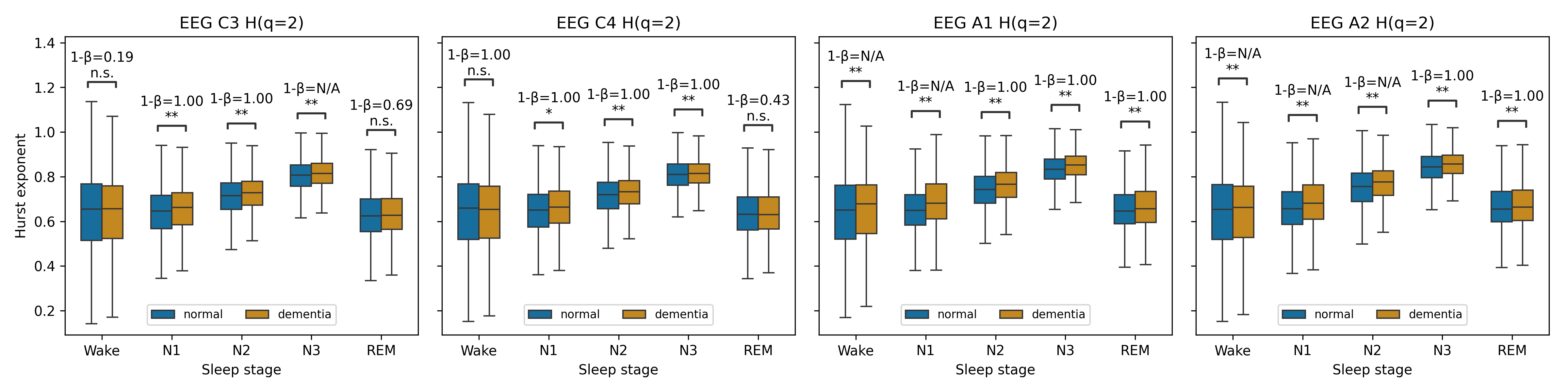}}
\caption{
    Distributions of Hurst exponents obtained from MFDFA ($q=2$) for each EEG electrode, stratified by the five sleep stages. Boxplots represent the median exponent value across subjects, with whiskers and boxes depicting the interquartile range (IQR). Statistical comparisons between cognitive groups were performed using a non-parametric permutation test. Significance levels are denoted as follows: $ns: p > 0.05$; $*: 0.01 < p \leqslant 0.05$, and $**: 0.001 < p \leqslant 0.01$. The notation $1-\beta = N/A$ indicates comparisons where statistical power could not be calculated due to insufficient sample size or zero variance.
    }
\label{fig:mfdfa}
\end{figure*}

\subsection{Multifractal Detrended Fluctuation Analysis (MFDFA)}\label{sec:mfdfa}

To characterize the complex, non-stationary dynamics of sleep EEG, we employed Multifractal Detrended Fluctuation Analysis (MFDFA)~\cite{MFDFA}. Unlike standard Detrended Fluctuation Analysis (DFA)~\cite{DFA_2}, which assumes a monofractal structure defined by a single scaling exponent, MFDFA accounts for the inherent heterogeneity of neural signals by identifying multiple scaling exponents that govern fluctuations at varying magnitudes.

In the context of neurodegenerative pathology, MFDFA serves as a robust proxy for signal criticality. Previous studies indicate that disease states, specifically Alzheimer’s disease and related dementias, are associated with a significant narrowing of the multifractal spectrum~\cite{10.3389/fnhum.2023.1155194}. This reduction in spectral width represents a loss of physiological complexity and a breakdown in the brain's capacity to maintain flexible, long-range temporal correlations across multiple time scales. By quantifying these shifts in the singularity spectrum, we aimed to evaluate the predictive utility of EEG criticality measures in identifying early-stage cognitive decline.

The MFDFA procedure for each 30-second EEG epoch was implemented in four primary stages~\cite{MFDFA}:
\begin{enumerate}
    \item \textbf{Integration:} The original time series $x(i)$ is transformed into an integrated profile $Y(i) = \sum_{k=1}^{i} [x(k) - \langle x \rangle]$ to enhance the signal-to-noise ratio and uncover scaling properties.
    \item \textbf{Segmentation:} The integrated profile is partitioned into $N_n = \text{int}(N/n)$ non-overlapping segments of equal length $n$.
    \item \textbf{Local Detrending:} For each segment, a local polynomial trend $y_{\nu,n}$ (of order $k=1$) is calculated via least-squares fitting and subtracted from the integrated profile to remove non-stationary ``drifts.''
    \item \textbf{Fluctuation Function Calculation:} The $q$-th order fluctuation function is computed as:
    \[ F_q(n) = \left\{ \frac{1}{2N_n} \sum_{\nu=1}^{2N_n} [F^2(\nu,n)]^{q/2} \right\}^{1/q} \]
    where the order $q$ determines the sensitivity to fluctuation magnitudes. \textbf{Positive $q$} values amplify the contribution of segments with large-scale fluctuations, while \textbf{negative $q$} values emphasize segments with small-scale fluctuations.
\end{enumerate}

The output is characterized by the generalized Hurst exponent $H(q)$, representing the slope of $\log F_q(n)$ vs. $\log n$. From this, the multifractal singularity spectrum $f(\alpha)$ is derived via the Legendre transform. The \textbf{spectral width} ($\Delta \alpha = \alpha_{max} - \alpha_{min}$) serves as a primary metric for signal complexity, where a broad spectrum indicates high multifractality (physiological health) and a narrow spectrum suggests a transition toward monofractal simplicity (pathological decline).

\subsection{Supervised UMAP on MFDFA Features}\label{sec:umap}

To visualize the manifold structure of the MFDFA features and assess their discriminative power, we applied Supervised Uniform Manifold Approximation and Projection (UMAP)~\cite{mcinnes2018umap}. Unlike standard unsupervised UMAP, which constructs a low-dimensional embedding based solely on topological data structure, the supervised variant incorporates categorical labels—specifically dementia status versus normal cognition—into the manifold learning process. These data were derived from the sleep polysomnography subset~\cite{spira2008sleep} of the Study of Osteoporotic Fractures (SOF) database~\cite{cummings1990appendicular}, hosted by the National Sleep Research Resource (NSRR)~\cite{zhang2018national}.

Mathematically, this introduces a categorical constraint during the optimization of the fuzzy simplicial set, penalizing edges that connect data points of different classes. This approach forces the embedding to prioritize the separation of cognitive groups while preserving the local and global relationships of the multifractal dynamics, thereby highlighting the ability of MFDFA features to capture neurophysiological differences associated with cognitive decline.

\section{Results}\label{sec:results}

MFDFA analysis of sleep EEG dynamics revealed distinct differences in signal criticality between the cognitively normal group and those progressing toward dementia. As illustrated in Fig.~\ref{fig:mfdfa}, the Hurst exponent $H(2)$ exhibited significant variability across sleep stages, a finding consistent across all four electrode locations. The most pronounced disparities occurred during NREM stages N1, N2, and N3, where the dementia group demonstrated values significantly closer to criticality at most electrode sites ($p \le 0.01$, non-parametric permutation test). While no significant median differences were detected at $C3$ and $C4$ during Wake and REM stages, significant differences emerged at the $A1$ and $A2$ electrodes ($p \le 0.01$). Statistical power ($1-\beta$) approached $1.0$ for the most significant comparisons, indicating a high likelihood of correctly rejecting the null hypothesis. However, some comparisons are noted as N/A where insufficient sample size or zero variance precluded power calculation.

\begin{figure*}[t]
\centerline{\includegraphics[width=\textwidth]{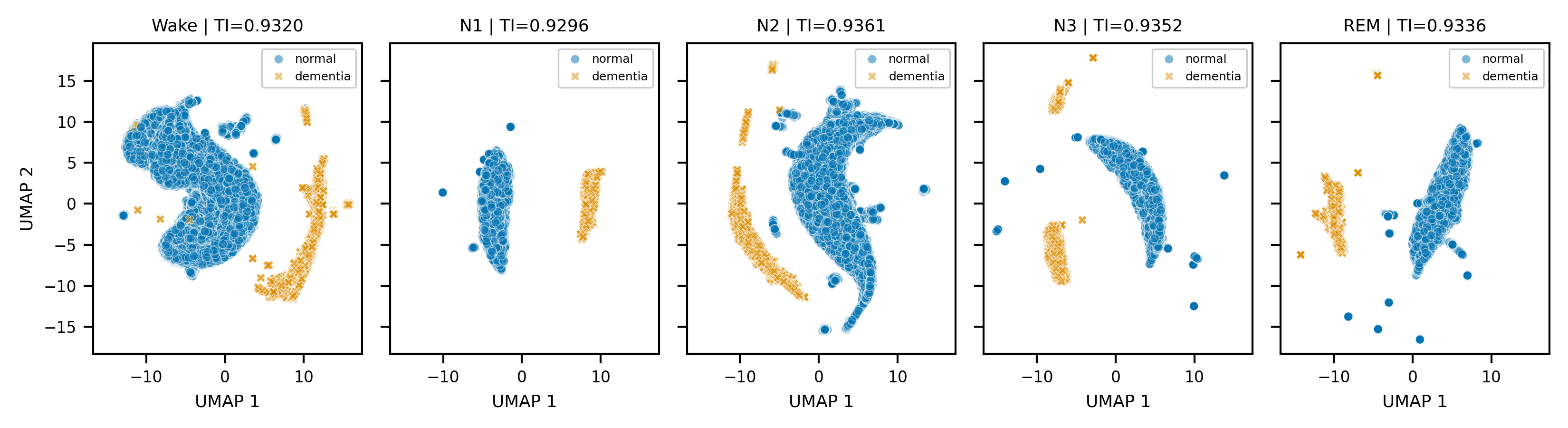}}
\caption{Supervised UMAP visualization of multifractal EEG dynamics. Two-dimensional embeddings of MFDFA features ($q = 2$) across sleep stages show robust topological separation between cognitively healthy controls and individuals with impairment (3MS score $< 78$). Trustworthiness Indices (TI)~\cite{Venna2001} near $1.0$ for each cluster plot confirm the high structural integrity of the high-to-low dimensional projection. Individual data points represent $30$-second epochs.}
\label{fig:umap}
\end{figure*}

To evaluate group separability based on multifractal features, Uniform Manifold Approximation and Projection (UMAP) was applied to MFDFA features from four EEG channels. The supervised UMAP projections (Fig.~\ref{fig:umap}) reveal a clear spatial divergence between cognitively normal participants and those progressing to dementia, validated by trustworthiness indices (TI)~\cite{Venna2001} that confirm cluster integrity. The most robust separation occurred during NREM (N1, N2, and N3) and REM sleep stages. Individuals progressing to cognitive impairment (3MS score $\le 78$) occupied a distinct region of the manifold space compared to the cognitively normal group. These results suggest that MFDFA-derived measures capture neurophysiologically meaningful differences that persist throughout the overnight sleep architecture.

\section{Conclusions}\label{sec:conclusions}

The findings of this study demonstrate that the criticality of sleep EEG signals, quantified through Multifractal Detrended Fluctuation Analysis (MFDFA), serves as a robust neurobiological marker for future cognitive decline. By analyzing the Study of Osteoporotic Fractures (SOF) cohort, we established that women who progressed to cognitive impairment within a five-year period exhibited distinct alterations in their Hurst exponent distributions compared to those who remained cognitively normal. 
The most substantial differences emerged during non-REM stages N2 and N3, where individuals progressing to dementia demonstrated values significantly closer to an optimally critical state across all electrode locations. This alignment with the Brain Criticality Hypothesis suggests that early impaired neural dynamics are defined by a scale-free complexity typically characteristic of awake states, thereby diverging from optimal deep-sleep patterns.
While the observed disparity during slow-wave sleep may hint at a disruption in the glymphatic clearance of neurotoxic proteins, such a mechanism remains speculative and warrants further direct investigation. The clarity of this divergence was further reinforced by supervised UMAP visualizations, which revealed a distinct spatial separation between groups across all sleep stages. This indicates that MFDFA-derived features effectively capture signal characteristics that facilitate the differentiation of participants based on their cognitive trajectories.

Despite these promising results, several limitations must be considered. The exclusive focus on an elderly female cohort limits the immediate generalizability of these findings to the broader population. Furthermore, the binary classification of cognitive status at a single follow-up point provides a snapshot rather than a continuous trajectory; future longitudinal tracking with multiple assessments would offer a more nuanced view of the temporal dynamics of criticality loss. While MFDFA is a powerful tool for characterizing signal complexity, it should be integrated with complementary measures of criticality in future studies to provide a more holistic view of neural state-space. Additionally, the observed overlap in Hurst exponent distributions suggests that while group-level differences are highly significant, individual-level diagnostics will likely require multivariate models that combine these dynamics with other physiological biomarkers.

Moving forward, exploring interventions that restore or maintain healthy neural dynamics—such as targeted sleep optimization strategies—represents a promising therapeutic frontier. If the approach towards criticality in deep sleep stages is a driver of cognitive erosion rather than a mere byproduct, then controlling these scale-free dynamics could offer a novel pathway for delaying dementia onset. Ultimately, this research underscores the potential of EEG criticality as a non-invasive, high-resolution window into the dementing brain. By identifying these signatures during the critical window of early impairment, we move closer to a future where automated, sleep-based screening tools can transform the early detection of neurodegeneration from a clinical challenge into a routine diagnostic reality.


\begin{thebibliography}{10}

\bibitem{livingston2024dementia}
G.~Livingston, J.~Huntley, K.~Y. Liu, S.~G. Costafreda, G.~Selb{\ae}k, S.~Alladi, D.~Ames, S.~Banerjee, A.~Burns, C.~Brayne, {\em et~al.}, ``Dementia prevention, intervention, and care: 2024 report of the lancet standing commission,'' {\em The Lancet}, vol.~404, no.~10452, pp.~572--628, 2024.

\bibitem{petersen2008mild}
R.~C. Petersen and S.~Negash, ``Mild cognitive impairment: an overview,'' {\em CNS Spectrums}, vol.~13, no.~1, pp.~45--53, 2008.

\bibitem{tuokko2003five}
H.~Tuokko, R.~Frerichs, J.~Graham, K.~Rockwood, B.~Kristjansson, J.~Fisk, H.~Bergman, A.~Kozma, and I.~McDowell, ``Five-year follow-up of cognitive impairment with no dementia,'' {\em Archives of Neurology}, vol.~60, no.~4, pp.~577--582, 2003.

\bibitem{xu2020sleep}
W.~Xu, C.-C. Tan, J.-J. Zou, X.-P. Cao, and L.~Tan, ``Sleep problems and risk of all-cause cognitive decline or dementia: an updated systematic review and meta-analysis,'' {\em Journal of Neurology, Neurosurgery \& Psychiatry}, vol.~91, no.~3, pp.~236--244, 2020.

\bibitem{sabia2021duration}
S.~Sabia, A.~Fayosse, J.~Dumurgier, V.~T. van Hees, C.~Paquet, A.~Sommerlad, M.~Kivim{\"a}ki, A.~Dugravot, and A.~Singh-Manoux, ``Association of sleep duration in middle and old age with incidence of dementia,'' {\em Nature Communications}, vol.~12, no.~1, p.~2289, 2021.

\bibitem{xie2013sleep}
L.~Xie, H.~Kang, Q.~Xu, M.~J. Chen, Y.~Liao, M.~Thiyagarajan, J.~O’Donnell, D.~J. Christensen, C.~Nicholson, J.~J. Iliff, {\em et~al.}, ``Sleep drives metabolite clearance from the adult brain,'' {\em Science}, vol.~342, no.~6156, pp.~373--377, 2013.

\bibitem{haghayegh2025predicting}
S.~Haghayegh, R.~Herzog, D.~A. Bennett, S.~Redline, K.~Yaffe, K.~L. Stone, A.~Ib{\'a}{\~n}ez, and K.~Hu, ``Predicting future risk of developing cognitive impairment using ambulatory sleep {EEG}: Integrating univariate analysis and multivariate information theory approach,'' {\em Journal of Alzheimer’s Disease}, p.~13872877251319742, 2025.

\bibitem{wilting201925crit}
J.~Wilting and V.~Priesemann, ``25 years of criticality in neuroscience—established results, open controversies, novel concepts,'' {\em Current Opinion in Neurobiology}, vol.~58, pp.~105--111, 2019.

\bibitem{tomen2019functional}
N.~Tomen, J.~M. Herrmann, and U.~Ernst, {\em The functional role of critical dynamics in neural systems}, vol.~11.
\newblock Springer, 2019.

\bibitem{blythe2014effect}
D.~A. Blythe, S.~Haufe, K.-R. M{\"u}ller, and V.~V. Nikulin, ``The effect of linear mixing in the {EEG} on hurst exponent estimation,'' {\em NeuroImage}, vol.~99, pp.~377--387, 2014.

\bibitem{zhang2018national}
G.-Q. Zhang, L.~Cui, R.~Mueller, S.~Tao, M.~Kim, M.~Rueschman, S.~Mariani, J.~L. Goodwin, and S.~Redline, ``The national sleep research resource: towards a sleep data commons,'' {\em Journal of the American Medical Informatics Association}, vol.~25, no.~10, pp.~1351--1358, 2018.

\bibitem{spira2008sleep}
A.~P. Spira, T.~Blackwell, K.~L. Stone, S.~Redline, J.~A. Cauley, S.~Ancoli-Israel, and K.~Yaffe, ``Sleep-disordered breathing and cognition in older women,'' {\em Journal of the American Geriatrics Society}, vol.~56, no.~1, pp.~45--50, 2008.

\bibitem{cummings1990appendicular}
S.~R. Cummings, D.~M. Black, M.~C. Nevitt, W.~S. Browner, J.~A. Cauley, H.~K. Genant, D.~Mascioli, J.~C. Scott, and J.~K. Seeley, ``Appendicular bone density and age predict hip fracture in women,'' {\em JAMA}, vol.~263, no.~5, pp.~665--668, 1990.

\bibitem{folstein1975mini}
M.~F. Folstein, S.~E. Folstein, and P.~R. McHugh, ``{`MINI-MENTAL STATE'}: A practical method for grading the cognitive state of patients for the clinician,'' {\em Journal of Psychiatric Research}, vol.~12, no.~3, pp.~189--198, 1975.

\bibitem{teng1987modified}
E.~Teng and H.~Chui, ``The modified mini-mental state {(3MS)} examination,'' {\em The Journal of Clinical Psychiatry}, vol.~48, p.~314—318, August 1987.

\bibitem{iber2007aasm}
C.~Iber, ``The {AASM} manual for the scoring of sleep and associated events: rules, terminology, and technical specification,'' {\em American Academy of Sleep Medicine}, 2007.

\bibitem{MFDFA}
J.~W. Kantelhardt, S.~A. Zschiegner, E.~Koscielny-Bunde, S.~Havlin, A.~Bunde, and H.~E. Stanley, ``Multifractal detrended fluctuation analysis of nonstationary time series,'' {\em Physica A: Statistical Mechanics and its Applications}, vol.~316, no.~1-4, pp.~87--114, 2002.

\bibitem{DFA_2}
C.-K. Peng, S.~Havlin, H.~E. Stanley, and A.~L. Goldberger, ``Quantification of scaling exponents and crossover phenomena in nonstationary heartbeat time series,'' {\em Chaos: An Interdisciplinary Journal of Nonlinear Science}, vol.~5, no.~1, pp.~82--87, 1995.

\bibitem{10.3389/fnhum.2023.1155194}
T.~M. Rutkowski, M.~S. Abe, T.~Komendzinski, H.~Sugimoto, S.~Narebski, and M.~Otake-Matsuura, ``Machine learning approach for early onset dementia neurobiomarker using {EEG} network topology features,'' {\em Frontiers in Human Neuroscience}, vol.~17, 2023.

\bibitem{mcinnes2018umap}
L.~McInnes, J.~Healy, and J.~Melville, ``{UMAP:} uniform manifold approximation and projection for dimension reduction,'' {\em arXiv preprint arXiv:1802.03426}, 2018.

\bibitem{Venna2001}
J.~Venna and S.~Kaski, ``Neighborhood preservation in nonlinear projection methods,'' in {\em Proceedings of the Symposium on Self-Organizing Maps}, pp.~273--280, 2001.

\end{thebibliography}

\addtolength{\textheight}{-12cm}   

\end{document}